\newcounter{myctr}

\documentclass{ws-acs}
\usepackage[T1]{fontenc} 
\usepackage{pgf} 
\usepackage{varioref} 
\usepackage{cite}

\begin{document}

\markboth{Huet S., Deffuant G.}{Openness Leads To Opinion Stability and Narrowness to Volatility}

%
\catchline{}{}{}{}{}
%

\title{OPENNESS LEADS TO OPINION STABILITY AND NARROWNESS TO VOLATILITY}

\author{HUET S.}
\address{Cemagref, Laboratoire d'Ingénierie des Systèmes Complexes\\
63172 Aubière, France\footnote{Cemagref, Laboratoire d'Ingénierie des Systèmes Complexes, BP 50085, 63172 F-Aubière}\\
sylvie.huet@cemagref.fr}

\author{DEFFUANT G.}

\address{Cemagref, Laboratoire d'Ingénierie des Systèmes Complexes\\
63172 Aubière, France\\
guillaume.deffuant@cemagref.cf}

\maketitle

\maketitle

\begin{abstract}
We propose a new opinion dynamic model based on the experiments and results of Wood et al (1996). We consider pairs of individuals discussing on two attitudinal dimensions, and we suppose that one dimension is important, the other secondary. The dynamics are mainly ruled by the level of agreement on the main dimension. If two individuals are close on the main dimension, then they attract each other on the main and on the secondary dimensions, whatever their disagreement on the secondary dimension. If they are far from each other on the main dimension, then too much proximity on the secondary dimension is uncomfortable, and generates rejection on this dimension. The proximity is defined by comparing the opinion distance with a threshold called attraction threshold on the main dimension and rejection threshold on the secondary dimension. With such dynamics, a population with opinions initially uniformly drawn evolves to a set of clusters, inside which secondary opinions fluctuate more or less depending on threshold values. We observe that a low attraction threshold favours fluctuations on the secondary dimension, especially when the rejection threshold is high. The opinion evolutions of the model can be related to some stylised facts.	
\end{abstract}

\section{Introduction}
\label{sec:Introduction}
Teenagers often tend to adopt any opinion expressed by their group or by their rock star idol, and reject any opinion expressed by their parents (or other representatives of the previous generation). Such unconditional influence and rejection mechanisms have been exhibited in several experiments [10] [24] [25]. Sociopsychological theories propose a variety of conceptual frameworks to interpret these observations: at individual level [2], [14], [26], [28] and at group level [29], [30] [3]. They generally claim influence is based on two complementary processes: attraction and rejection. Nevertheless, rejection mechanisms are still poorly understood.

Social simulation can provide complementary evidences on how mechanisms at individual level generate collective regularities in large populations, whereas experiments of social psychology are limited to small groups. In this perspective, the effect of the opinion or attitude [9] attraction  has been widely studied [5, 7, 8, 12, 13, 21]. Fewer works are dedicated to both attraction and rejection mechanisms [18-20], [27], [31, 32]. We have recently proposed a model inspired from the socio-psychological theories [17] (dissonance [10], the social judgement [28] and the self-categorization [30]), which couples attraction and rejection in a multidimensional approach as in [1], [22], [6, 11, 16, 19, 20, 31, 32]. We observe that this model generally leads to more consensus than when attraction is the only mechanism. In this paper, we propose a refinement of our previous model which is directly inspired from Wood et al. experiment [33]. 

Wood et al. describe their main results as follows: in a first study, \textit{participants who considered a majority group relevant to their own self-definitions (but not those who judged it irrelevant), on learning that the group held a counter-attitudinal position, shifted their attitudes to agree with the source. In a second study, recipients who judged a minority group (negatively) self-relevant, on learning that the group held a similar attitude to their own, shifted their attitudes to diverge from the source. These shifts in attitudes were based on participants' interpretations of the attitude issues}. The authors suggest that these attitude shifts reflect normative pressures to align with valued groups and to differentiate from derogated groups. Globally, these experiments show that:

\begin{itemize}
	\item one can be attracted by a far opinion (counter-attitudinal), if this opinion comes from a group (which can be majority but not necessary) sharing other fundamental values (relevant to their own self definition),
	\item one tends to shift from a close opinion expressed by a group (minority or not) with significant differences about fundamental values (negatively self relevant).
\end{itemize}

We propose a new model derived from [34], which aims at reproducing these observations. However, modelling implies interpretations and simplifications. First, we study group formation from initially scattered opinions, whereas the experiments took place in the context of already existing groups. Thus, in our model, we consider pair interactions only (we do not consider the group directly), and we reinterpret the experiment by considering two individuals who can share or not fundamental values, and influence each other on less important (secondary) opinions. Second, we consider only two attitudinal dimensions, one is for fundamental values (called main dimension), the other more secondary opinions (called secondary dimension). On each dimension, we suppose that the opinions can take continuous values, between -1 and +1. Then, we introduce influence dynamics: When the individuals are close to each other on the main dimension, they tend to attract each other on the secondary dimension, even if their disagreement is strong. When they disagree on the fundamental dimension, and are close on the secondary, then they tend to reject each other's secondary opinions. To model a disagreement, we compare the distance between opinions with a threshold (called attraction threshold on the main dimension, and rejection threshold on the secondary dimension). 

We consider populations of individuals with opinions initially drawn at random, and simulate their evolutions with these mechanisms, for different values of the parameters (mainly the attraction and rejection thresholds). The stationary state shows a set of clusters, with fluctuations on the secondary dimension. These fluctuations are lower or absent when the attraction threshold is high. They cover the whole length of the opinion axis when both the attraction and rejection thresholds are low. Moreover, a high rejection threshold tends to generate some clusters with extreme opinions on the secondary dimension (polarization). We propose a theoretical relation between the thresholds values which allows us to predict large fluctuations and polarisation.

This article firstly presents the proposed model. A second section describes typical opinion evolutions of the population. These first observations lead us to formulate our main hypothesis on the parameter values leading to fluctuations or polarization on the secondary dimension. Then we propose a more complete experiment design to check this hypothesis. A last section discusses and concludes this work.

\section{The dynamic model of interacting individuals}
\label{sec:DynamicInteractingIndiv}

We consider a population of \textit{N} individuals. The model includes three parameters: $u_{m}$ and $u_{s}$, respectively the attraction and rejection thresholds, and $\mu$ ruling the intensity of influence at each meeting (comprised between 0 and 0.5). 

An individual has 2 opinions $x_{m}$ (on the main dimension) and $x_{s}$ (on the secondary dimension) taking real values between -1 and +1. 

During an iteration, a couple of individuals \textit{X} and \textit{Y} is randomly chosen and can influence each other. The algorithm is the following:

\begin{itemize}
\item	\begin{verbatim} Choose randomly a couple (X,Y) of individuals in the
population; \end{verbatim}
\item	\begin{verbatim} X and Y change their opinions at the same time, according to
 the influence function. \end{verbatim}
\end{itemize}

We present the calculation of the influence of \textit{Y} on \textit{X} (of course the influence of \textit{X} on \textit{Y} is found by inverting \textit{X} and \textit{Y}). Let ($x_{m}$, $x_{s}$) and ($y_{m}$, $y_{s}$) be the opinions of \textit{X} and \textit{Y} respectively. We first consider the main opinion dimension:

\begin{itemize}
	\item If $\left|x_{m} - y_{m}\right| \leq u_{m}$, individual \textit{X} agrees with \textit{Y} on the main dimension. Both attitudes of \textit{X} are going to get closer to those of \textit{Y}, proportionally to the attitudinal distance on each dimension:
 
\begin{equation}
x_{m}\left(t+1\right) = x_{m}\left(t\right) + \mu\left(y_{m}\left(t\right)-x_{m}\left(t\right)\right)
\end{equation}  
\begin{equation}
x_{s}\left(t+1\right) = x_{s}\left(t\right) + \mu\left(y_{s}\left(t\right)-x_{s}\left(t\right)\right)
\end{equation}

Indeed, whatever the agreement level on the secondary dimension is, \textit{X} is going to be globally closer to \textit{Y}. If it was already close, it gets closer. Even if it was far on this dimension, following the spirit of the experimental observations, the closeness on the main dimension leads to get closer in the secondary dimension.

	\item If $\left|x_{m} - y_{m}\right| > u_{m}$ , individual \textit{X} disagrees with \textit{Y} on the main dimension and if $\left|x_{s} - y_{s}\right| \leq u_{s}$: Individual \textit{X} feels it is too close to \textit{Y} on the secondary dimension, because of their disagreement on the main dimension. To solve the conflicting situation, \textit{X} moves away from \textit{Y} on this dimension. The attitude change is proportional to the distance to reach the rejection threshold: 

if $\left(x_{s} - y_{s}\right) < 0$ then 
\begin{equation}
x_{s}\left(t+1\right) = x_{s}\left(t\right) - \mu\left\{u_{s} - \left(y_{s}\left(t\right)-x_{s}\left(t\right)\right)\right\}
\end{equation} 
else 
\begin{equation}
x_{s}\left(t+1\right) = x_{s}\left(t\right) + \mu\left\{u_{s} + \left(y_{s}\left(t\right)-x_{s}\left(t\right)\right)\right\}
\end{equation}  
	
\end{itemize}

In the other cases \textit{X} is not modified by \textit{Y}.

Moreover, we confine the attitude in the interval [-1, +1]: if $\left|x_{i}\right|>1$\ then $x_{i} :=sign(x_{i})$\ where \textit{sign()} is a function which returns -1 if its argument is strictly negative, +1 otherwise.

The attitude of \textit{Y} is calculated is the same way considering the situation of the meeting with \textit{X}.

\section{Typical evolutions of the population}
\label{sec:TypEvolPop}

As known for the classical bounded confidence model and its extensions, the most significant parameter is the threshold limiting the confidence (sometimes called uncertainty). Here the attraction and rejection thresholds $u_{m}$ and $u_{s}$ are the main parameters. We consider three situations, for which we study the population opinion evolutions: 

\begin{itemize}
	\item $u_{m}$ = $u_{s}$;
	\item $u_{m}$ > $u_{s}$;
	\item	$u_{m}$ < $u_{s}$.
\end{itemize}

We firstly describe the initialisation parameters and the experimental design. Then, each possible typical evolution of the population is shown by several two-dimensional graphs representing each the two-attitude space. Each graph draws at different times the attitudes of individuals of the population. Globally, a figure, composed of a set of graphs, shows the evolution of the attitudes on time for a run of the model. Finally, we formulate some hypothesis about the global behaviour of the model.

\subsection{Model initialisation and experiments}
\label{sec:Model initialisation and experiments}

In all simulations, the following values are fixed:

\begin{itemize}
	\item All individuals have the same speed of attitude change $\mu$ = 0.5 on the two attitudinal dimensions;
	\item The main attitude dimension is the horizontal one (the secondary dimension is the vertical one);
	\item The size of the population equal to 1500 individuals.
\end{itemize}

\begin{figure}[h]
	\centering
		\includegraphics[scale=0.35]{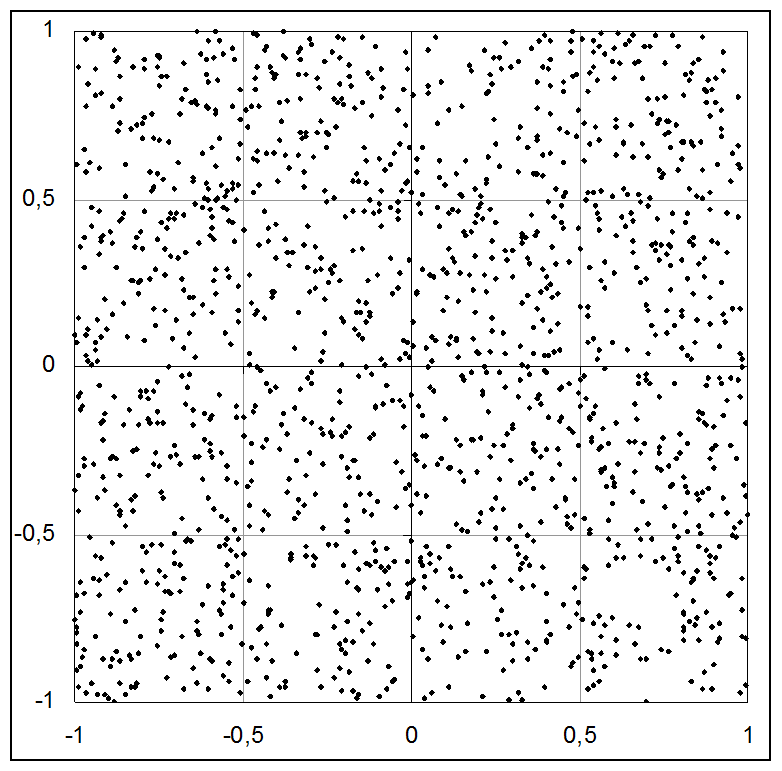}
		\caption{Initial state for of a 1500-individual population}
	\label{fig1}
\end{figure}

The attitudes $x_{i}$ on each dimension are initialised following a Uniform law and comprised between -1 and 1. Such an initialization is presented in figure \ref{fig1} for which each axe represents one attitudinal dimension varying from -1 to 1. Thus, each point corresponds to the coordinates of an individual's attitudes.

We run the model until it reaches a stationary state. 

\subsection{Attraction and rejection thresholds are equal}
\label{sec:EqualThresholds}

\begin{figure}[h]
	\centering
		\includegraphics[scale=0.6]{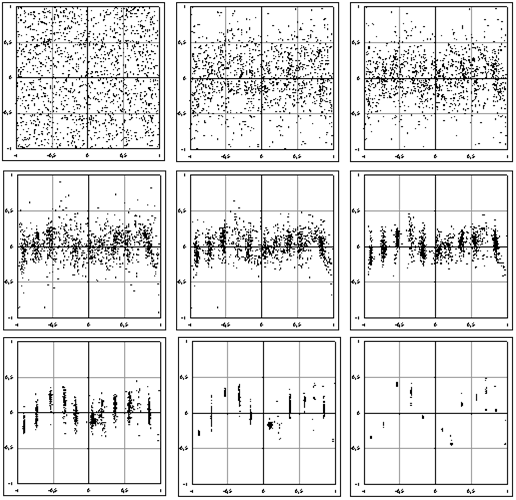}
	\caption{Population evolution (from upper left to lower right) at step number 0, 30000, 37500, 52500, 67500, 90000, 105000, 172500, 375000, of a 1500-individual population for $u_{m}$ = 0.1, $u_{s}$ = 0.1. Main dimension is horizontal.}\label{fig2}
\end{figure}

Figure \ref{fig2} shows an example of the evolution of the opinions when the attraction and rejection thresholds are equal. We firstly observe an attraction between individuals on the secondary dimension (y-axis) due to their proximity on the main dimension (x-axis). Then, clusters appear on the main dimension. While groups are forming close to the middle of the secondary dimension, they begin to reject each other. This is due to the disagreement between groups on the main dimension, which lead to rejection between close opinions on the secondary dimension. Thus, most of the groups polarize on the secondary dimension. This means that the average y-attitude value of each group increases in absolute value. We finally observe that the number of clusters is defined by the attraction threshold $u_{m}$. This value defines the minimum distance between two groups preventing them to merge.

\subsection{Attraction threshold larger than rejection threshold}
\label{sec:AttractionLarger}

\begin{figure}[h]
	\centering
		\includegraphics[scale=0.45]{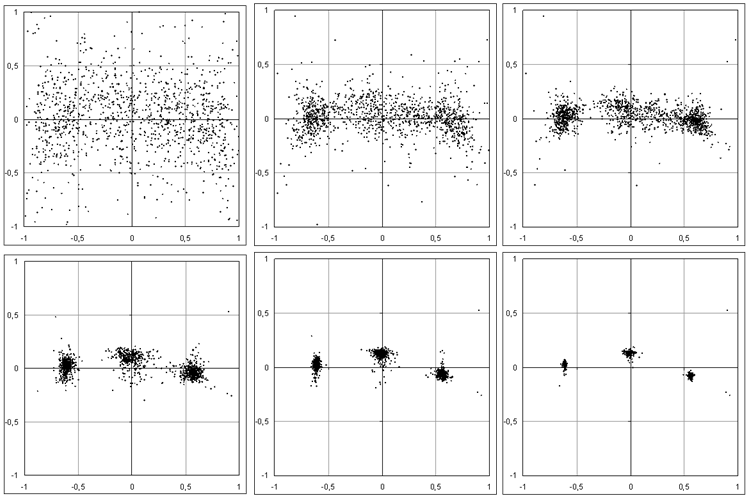}
	\caption{Population evolution (from upper left to lower right) at step number 11250, 22500, 37500, 45000, 52500, 67500, of a 1500-individual population for $u_{m}$ = 0.3, $u_{s}$ = 0.1. Main dimension is horizontal.}\label{fig3}
\end{figure}

Figure \ref{fig3}, shows an example of simulation with $u_{m}$ = 0.3, $u_{s}$ = 0.1. Again, we observe the attraction on the secondary dimension first, and then, clusters appear on the main dimension, with a gravity centre close to 0 on the secondary dimension. As previously, the clusters tend to reject each other on the secondary dimension, but this effect is lower because the rejection threshold is lower.

As in the previous section, the number of large clusters is determined by the attraction threshold. We also observe some minor clusters on the border of the main attitude space.

\subsection{The attraction threshold is smaller than the rejection threshold}
\label{sec:AttractionSmaller}

\begin{figure}[h]
	\centering
		\includegraphics[scale=0.45]{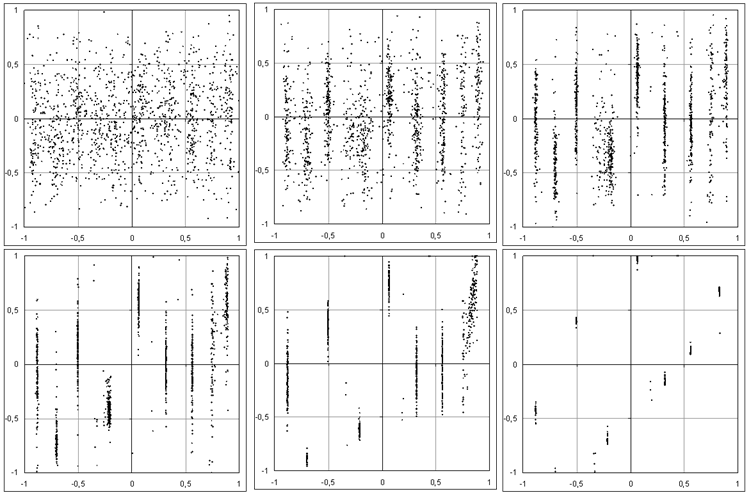}
	\caption{Population evolution (from upper left to lower right) at time step number 30000, 90000, 127500, 168750, 225000, 375000, for $u_{m}$ = 0.1, $u_{s}$= 0.3.}\label{fig4}
\end{figure}

Figure \ref{fig4} shows the evolution of the population for $u_{m}$ = 0.1, $u_{s}$ = 0.3. The pattern of opinion evolution is slightly different. The initial attraction on the secondary dimension is very weak. The polarization on this dimension begins before the clusters have been really formed. Indeed, as the rejection threshold is high, the conditions for rejection are met more often. Moreover, the clusters must have a higher distance on the secondary dimension to reach some stability. The stability is not complete, because some fluctuations remain in the clusters, on the secondary dimension, it is not possible to keep a distance of 0.3 between 10 clusters on a distance of 2 overall. Hence some rejection continues to take place between the groups at the stationary state.

\begin{figure}[h]
	\centering
		\includegraphics[scale=0.45]{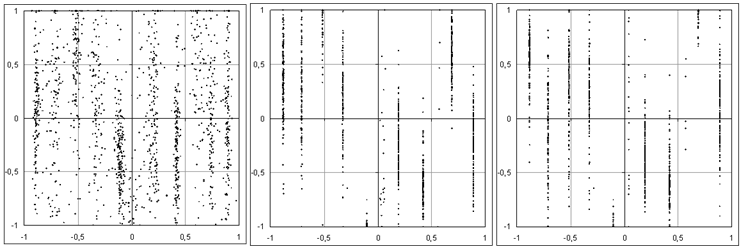}
	\caption{Population evolution at time step number 90000, 187500, 375000, $u_{m}$ = 0.1, $u_{s}$ = 0.5 }\label{fig5}
\end{figure}

\begin{figure}[h]
	\centering
		\includegraphics[scale=0.45]{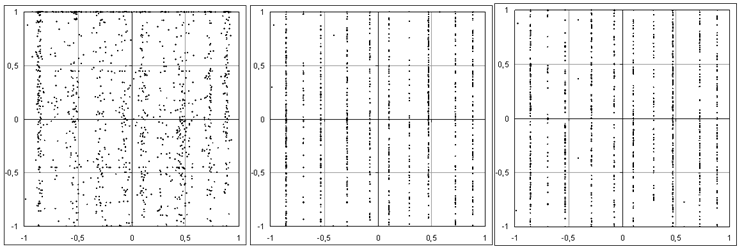}
	\caption{Population evolution at time step number 90000, 187500, 375000, for $u_{m}$ = 0.1, $u_{s}$ = 1.1.}\label{fig6}
\end{figure}

This effect increases when the rejection threshold increases (see Fig. \ref{fig5} and \ref{fig6}). The fluctuations on the secondary dimension reach almost the whole dimension space on figure \ref{fig6}, for a rejection threshold $u_{s}$ = 0.5. 

\begin{figure}[h]
	\centering
		\includegraphics[width=5cm]{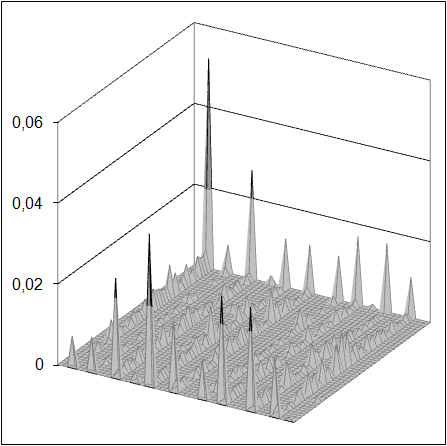}\hfill
	  \includegraphics[width=5cm]{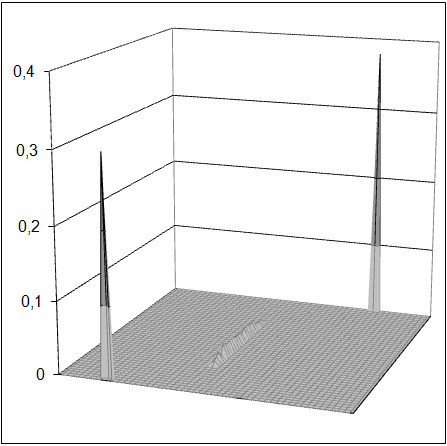}
	\caption{Density of individuals on the attitude space at the equilibrium for a 1000-individual population:  for $u_{m}$ = 0.1, $u_{s}$ = 1.1 on the left; for $u_{m}$ = 0.3, $u_{s}$ = 1.5 on the right}\label{fig7}
\end{figure}

For $u_{m}$ = 0.1, $u_{s}$ = 1.1 (Fig. \ref{fig6}), a larger density of individuals appear on the borders of the attitude space as shown on figure \ref{fig7} on the left. On the right with $u_{m}$ = 0.3, $u_{s}$ = 1.5 this tendency is enhanced with two stable extreme clusters, and a central cluster where secondary attitudes fluctuate. The three groups include similar number of individuals.

\subsection{Hypothesis about the global behaviour of the model}
\label{sec:Hypothesis about the global behaviour of the model}

In the area of our experimental plan, and following this first observation of the model evolutions, we make the following hypothesis on the global dynamics.

\begin{enumerate}
	\item The final number of large clusters is approximately $(1/u_{m})$, because this number is ruled by the dynamics on the main dimension, the classical bounded confidence model [8]. Indeed, on the secondary dimension y, there cannot be two clusters on the same vertical line, because these clusters tend to merge whatever their distance. Let's notice that 1 corresponds to the mid width of an attitudinal dimension (2 is the total width of an attitudinal dimension).
	\item On the secondary dimension, if the space is sufficient to get clusters distant from each other of more than the rejection threshold $u_{s}$, the final state is static. Otherwise, there are constant fluctuations due to the rejection, in the stationary state. Regarding the approximation of the total number of clusters, the maximum possible distance between two clusters on the secondary dimension is:
\begin{equation}
\delta=\frac{2}{\left(\left(1/u_{m})\right)-1\right)} 
\end{equation}	\end{enumerate}

Thus, if $u_{s}\leq\delta$ there is no fluctuations on the secondary dimension. But, when $u_{s}$ gets close to $\delta$, the clusters tend to occupy the whole length of the secondary dimension. For $u_{s}>\delta$ , the clusters are necessarily rejecting each other on the secondary dimension, because their distance is less than $u_{s}$. The opinions fluctuate more and more on the secondary dimension, when $u_{s}$ increases. 

Note that when $u_{m}$ is small, $\delta$ is close to 2$u_{m}$. In this case, one can summarise that fluctuations on the secondary dimension appear when the rejection threshold is about higher than twice the attraction threshold.

From the study of opinion evolutions, we can also claim that, in the no fluctuation zone, larger is the rejection threshold, higher is the polarisation. Indeed, after the initial attraction to the neutral position on the secondary dimension, groups tend to increase in absolute value their average opinion until they reach one avoiding the conflict with the other groups.

We are now going to check our hypothesis with more systematic experiments.

\section{Systematic experiments}
\label{sec:Systematic experiments}

\subsection{Initialisation of the model and experiments}
\label{sec:Initialisation of the model and experiments}

\begin{figure}[h]
	\centering
		\includegraphics[scale=0.65]{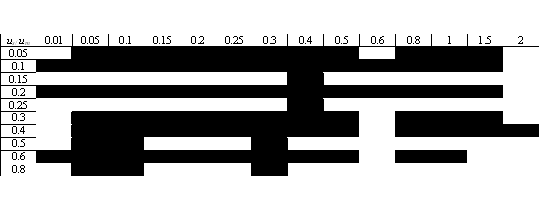}
	\caption{Tested values (in black) for $u_{m}$ (columns) and $u_{s}$ (lines) with a population of 1000 individuals}\label{fig8}
\end{figure}

The initialization is the same as the one presented in 3.1. The model runs during 40.000.000 iterations. This is always sufficient to attain the stationary state. In several experimental designs, we vary systematically the values of the attraction and rejection threshold. We study the number of clusters (for two population sizes: 1000 and 7500 individuals), the presence of fluctuations and the polarisation for a population of 1000 individuals (see figure \ref{fig8}). 

The presented results are the average, and sometimes the maximum and the minimum, of the measured values on 20 replicas run for each set of parameter values. The final number of clusters is computed via a classical algorithm searching for the chains of individuals separated by a maximum distance (the chosen distance for this model is min($u_{m}$, $u_{s}$)).

\subsection{Final number of clusters}
\label{sec:Final number of clusters}

\begin{figure}[h]
	\centering
		\includegraphics[width=6.2cm]{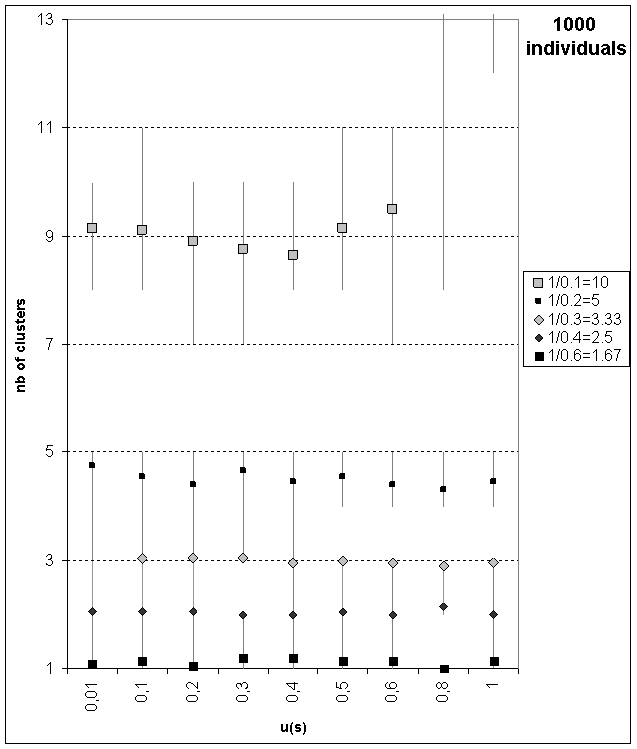}\hfill
		\includegraphics[width=6.2cm]{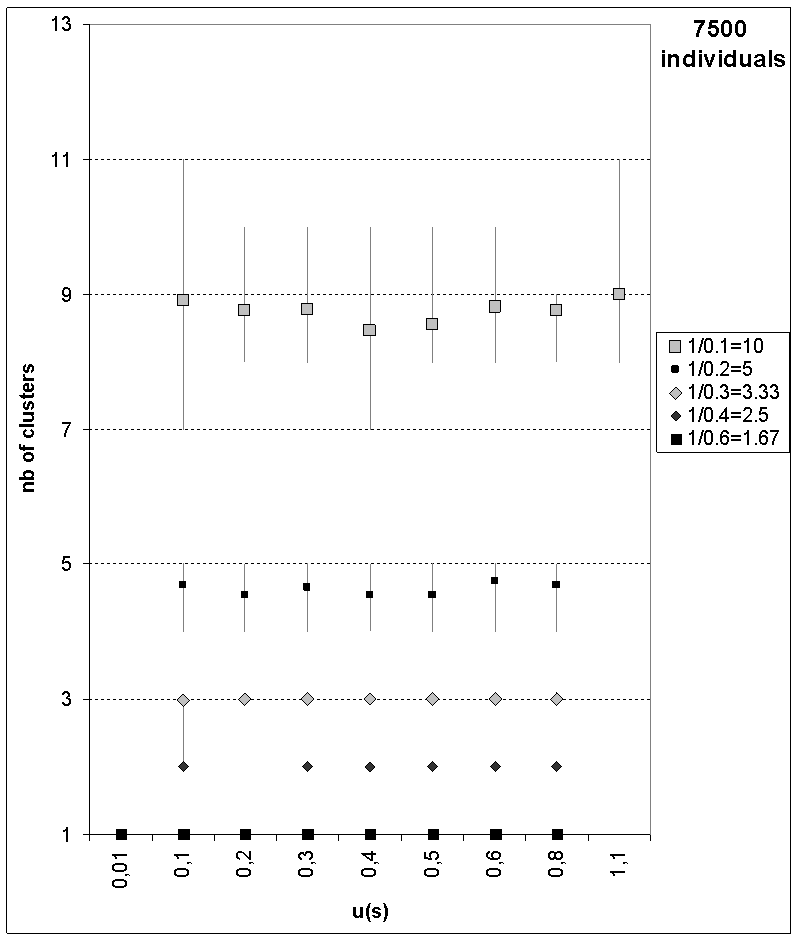}\caption{(on the left) Average final cluster numbers, each representing at least 1\% of the population, for various values of $u_{m}$   and $u_{s}$  for  $u_{m}$  equal to (from the bottom to the top: 0.6, 0.4, 0.3, 0.2, 0.1) for 1000 individuals (on the left) and 7500 individuals (b on the right). The legend indicates the number of clusters which can be calculated from the simulated values of $u_{m}$. One can see the results of these calculations are quite similar to the simulation results.}\label{fig9}
\end{figure}

Figure \ref{fig9} shows the counted average, maximum and minimum, number of clusters representing each more than 1\% of the population. 

Firstly we observe that the size of population doesn't change the final number of clusters. Indeed, the diagram for 1000 individuals (see figure \ref{fig9} on the left) is close to the diagram for a population of 7500 individuals (see figure \ref{fig9} on the right).

The most important to observe is that, for the both population sizes, the average number of major clusters approximatively corresponds to $(1/u_{m})$. Regarding the number of clusters containing more than  1\% of the population, the model has exactly the same behaviour as the classical bounded confidence model [8]. The exceptions are due to a default of the algorithm counting the cluster for small population. Indeed, when $u_{m}$ is lower and lower, the algorithm becomes more and more inefficient especially when the fluctuations on the secondary dimension are large.

\subsection{Presence of fluctuations and polarization}
\label{sec:Presence of fluctuations and polarization}

\begin{figure}[h]
	\centering
		\includegraphics[width=6cm]{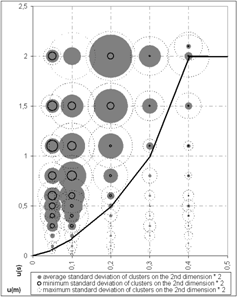}\hfill
		\includegraphics[width=6cm]{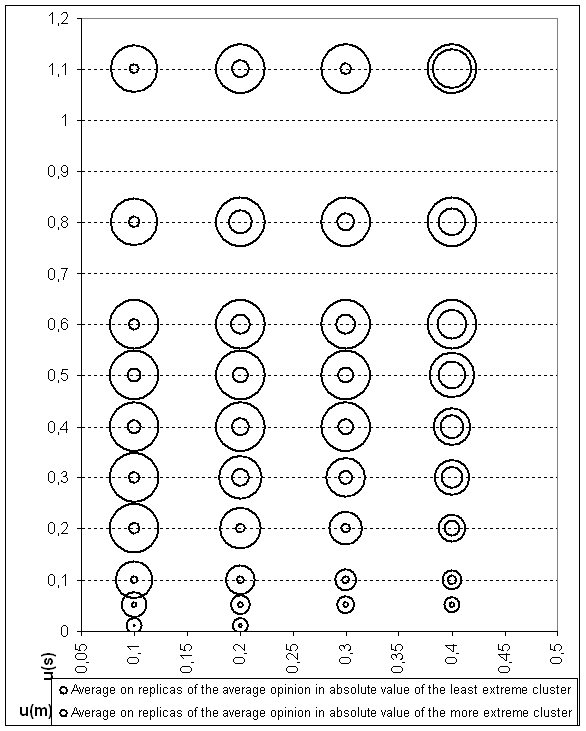}
	\caption{(a) Average, minimum and maximum on the 20 replicas of the standard deviation of cluster for the secondary  dimension and $\delta$ which is represented by the dark line (on the left); and (b) Average on replicas of the average attitude on the secondary dimension of the least and the more extreme cluster (on the right) for various values of $u_{m}$  and $u_{s}$  and a population of 1000 individuals.}\label{fig10}
\end{figure}

Figure \ref{fig10} shows the results of sets of simulations defined by the values of $u_{m}$ (x-axis) and  $u_{s}$ (y-axis).

For each replica, we calculate at the equilibrium state the opinion standard deviation  of each cluster. Then, we compute for the replica the average, minimum and maximum standard deviations on the clusters. We finally compute the average, minimum and maximum standard deviations on all the replicas for the considered couple of values $(u_m, u_s)$. Hence we get, for each couple $(u_m, u_s)$,  the average on the replicas of the average standard deviations on clusters (represented by a grey disc on figure \ref{fig10} left), the minimum on the replicas of the minimum standard deviations on clusters (represented by a dark circle on figure \ref{fig10} left), the maximum on the replicas of the maximum standard deviations on clusters (represented by a dotted circle on figure \ref{fig10} left). This indicator gives a rough idea of the width of the clusters on the secondary dimension. Indeed, we have seen in figures \ref{fig4} to \ref{fig6} that fluctuations enlarge the clusters on the secondary dimension. The dark line represents $\delta$. The figure \ref{fig10} on the left confirms that large fluctuations occur when $u_{s}>\delta$  since the width of the clusters increases in this zone. It also confirms that the fluctuations increase when $u_{s}$ increases. However, while our hypothesis claims that the final state is static for $u_{s}\leq\delta$, we observe on the graph some fluctuations, especially when $u_{m}$ is large.

On the right of the figure \ref{fig10}, we can observe the average attitude on replicas of the secondary dimension of the least and the most extreme clusters. One can see, as we have hypothesized, that the polarization increases when $u_{s}$ increases until it reaches a plateau. However, as previously noticed for fluctuations, we observe extreme polarization for some $u_{s}\leq \delta$ while our hypothesis predicts less polarization at this stage.

\begin{figure}[h]
	\centering
		\includegraphics[width=6cm]{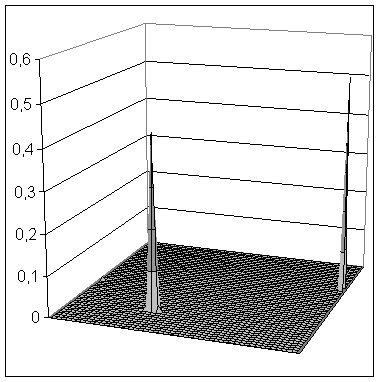}\hfill
		\includegraphics[width=6cm]{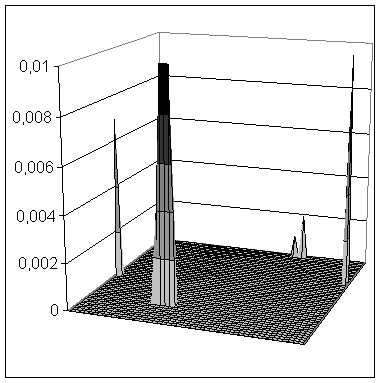}
	\caption{Density of individuals on the attitude space at the equilibrium for a 1000-individual population  for $u_{m}$ = 0.4, $u_{s}$ = 0.8: on the left at the normal size; zoomed on the lower part of the graph on the right to be able to observe the minority clusters.}\label{fig11}
\end{figure}

To sum-up, our hypothesis is confirmed for $u_{s} > \delta$ but not for $u_{s} \leq \delta$. Indeed, even the observed fluctuations for $u_{s}\leq \delta$ are on average significantly lower, the final states are not static for at least some replicas. How to explain these intra-group fluctuations? The dynamics on the main dimension is ruled by the classical bounded confidence model [8]. We know, from the study of this model done by [35] that some minor clusters appear sometimes between the major cluster and on the border of the attitude space. Their masses are around $3.10^{-4}$. The hypothesis we can make is that the presence of such minor cluster is responsible of these fluctuations that we observe for $u_{s}\leq \delta$. Indeed, even if it includes a single individual, a cluster can strongly modify the global configuration of the population in the dynamics we are studying.

\paragraph() 
This is illustrated by Figure \ref{fig11}, which shows a final state for $u_{m}$ = 0.4, $u_{s}$ = 0.8 on the left. On the right, there is a focus on the lower part of the diagram presented on the left. We can see on the right the presence of much smaller cluster. These small clusters reject the big ones, and increase the expected polarization. In other cases, their presence can generate fluctuations.

\begin{figure}[h]
	\centering
		\includegraphics[width=6cm]{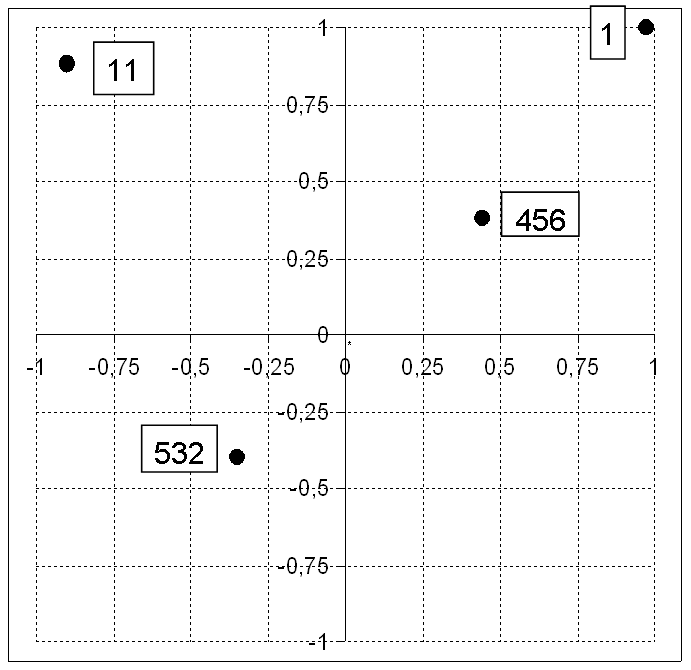}
	\caption{Clusters of individuals in their attitude space for a 1000-individual population after 7500000 iterations for $u_{m}$ = 0.4, $u_{s}$ = 0.5. Each dot represent a cluster and the labels indicate their size. The x-axis represents the main attitude, the y-axis the secondary attitude.}\label{fig12}
\end{figure}

In a closer analysis of the simulations, we found a second important difference with the general hypothesis which led to the computation of $\delta$: a theoretical stable final state can be very difficult to reach in practise, and require an extremely long time, during which fluctuations are observed. For instance, figure \ref{fig11} presents the opinion clusters for a 1000-individual population after 7500000 iteration for $u_{m}$ = 0.4, $u_{s}$ = 0.5. Each dot represents a cluster and the label indicates its size. We count 4 clusters on the figure. TIf we follow our first analysis, this configuration should be stable because $u_{s} \leq 2/(4-1)$ $\delta$. However, one can see on the figure that the two minor clusters are not stable because they are too close on the secondary dimension. In this configuration, the necessary space between the clusters on the $u_s$ axis should be obtained by the progressive drift of the big clusters under the influence of the small ones: the cluster having size 1 pushes the one of size 11, which in turn, is pushes the cluster of size 456, which itself pushes the cluster having a size of 532. The time required to get the stability is so long that it is difficult to observe it in practise. Therefore, in some configurations, the minor clusters can induce fluctuations that remain for such a long time that we did not observe the final stable state in our experiments. 

\begin{figure}[h]
	\centering
			\includegraphics[width=6cm]{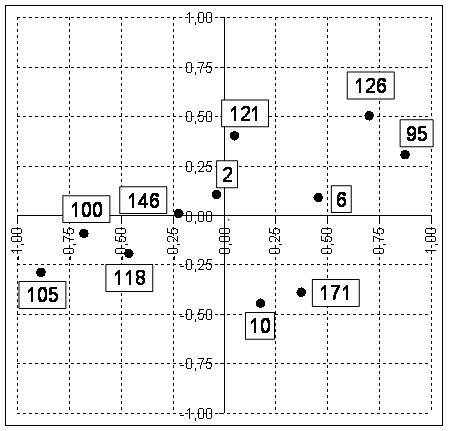}\hfill
		\includegraphics[width=6.5cm]{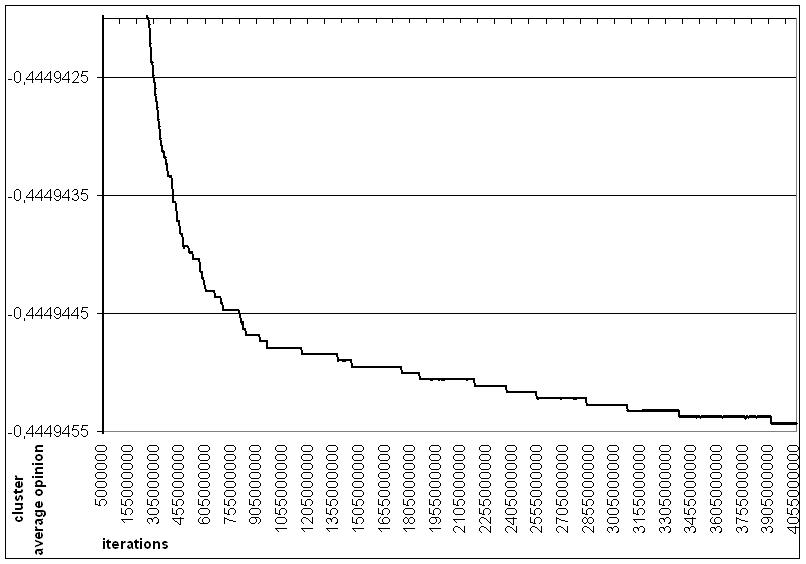}
	\caption{On the left: Clusters of individuals in their attitude space for a 1000-individual population after 4005000000 iterations for $u_{m}$ = 0.1, $u_{s}$ = 0.1. Each dot represent a cluster and the labels indicate their size. The main attitude is represented in abscissa while the secondary is on the y-axis. On the right: evolution of the average opinion on the least important dimension (y) of the more extreme negative cluster regarding this dimension (it has the size 10)}\label{fig13a}
\end{figure}

Figure \ref{fig13a} shows another example of this very long fluctuating transitory state for $u_{m}$ = 0.1, $u_{s}$ = 0.1. We notice on the right that after 4005000000 iterations, the clusters are far from occupying the whole space of the secondary dimension and remain in a conflicting situation. On the right, we see how the average opinion of the most extreme negative cluster slowly changes to find a stable position. It gives an idea of the time required to lead the stable state.  

\subsection{Size of the clusters}
\label{sec:Size of the clusters}

Figure \ref{fig12} shows on abscissa the tested values of the attraction and rejection thresholds. It presents the average size of the biggest and the smallest cluster (expressed in percentage of the population size). The graph also shows what would be the size of the clusters if this size is equal for all the clusters. This size is calculated considering the population size divided by the measured average number of clusters larger than 1\% of the population. All are expressed in percentage of the population.

We observe that, for a given value of $u_{m}$ and $u_{s}$, the clusters are approximately all of the same size in the population. Indeed the "identical for all clusters" size is always at most more or less 4\% of the size of the smallest and the largest cluster.

\begin{figure}[h]
	\centering
		\includegraphics[scale=1]{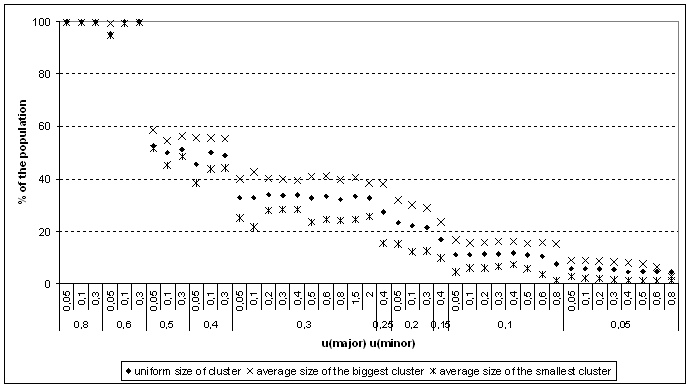}
	\caption{Average size of the biggest and the smallest cluster and size of clusters if all clusters have equal size (it means that the size of a cluster corresponds to the size of the population divided by the average measured number of clusters larger than 1\% of the population) for various values of $u_{m}$ and $u_{s}$ (all expressed in \% of the population composed from 1000 individuals).}\label{fig12}
\end{figure}

\section{Discussion and conclusion}
\label{sec:DiscussionAndConclusion}

The experimental results are in good accordance with our hypothesis. The number and the size of clusters are ruled by the bounded confidence dynamics on the main dimension. This behaviour is not modified by the population size, as for the bounded confidence model. The measures on fluctuations and polarisation confirmed our hypothesis for $u_{s}>\frac{2}{\left(\frac{1}{u_{m}}-1\right)}$ and disconfirmed, in a finite time, for the other values of $u_{s}$. While we predict static and low polarized clusters for these latter values, we observe quite highly polarized clusters and intra-group fluctuations. This is due to the presence of minor clusters which imply a very long continuous rejection making the stable state reachable in a very very long time, sometimes too long in practice. The model is ruled by the bounded confidence model on the main dimension and, for this model, minor clusters regularly appear [35][36] between the big clusters and on the border of the attitude space, as pointed out by [35][36].

Now, the main discussion is about the interpretative potential of this model. Do its typical opinion evolutions fit observed stylised facts? 
The first typical opinion evolution is obtained for $u_{s}\leq\frac{2}{\left(\frac{1}{u_{m}}-1\right)}$, meaning for these values, the simulations are mainly ruled by the attraction process (we suppose here that the width of the attitudinal dimension is 2 because the attitudes take values between -1 and +1). Individuals begin to discuss and they quickly agree on the secondary attitudinal dimension: they all join the mid position on this dimension. They act exactly as people who easily agree on details. Then, individuals form clusters on the main dimension. The most important aspects define each group as a unique entity. When the groups are sufficiently formed, individuals begin to reject each other on the secondary dimension due to their high distance on the main dimension. We get a behaviour which reminds the results of experiments: when they belong to different groups (defined on the main attitude), individuals having the same attitude on secondary aspects reject each other. It is also very close to the process of group formation and the increase of the cohesion described by Turner in 1984 [34].

For these opinion evolutions, we can also observe a great effect of small minorities. They almost always exist even if they generally represent less than one percent of the population. They maintain some intra-group fluctuations in the major groups due to their rejection for a very long time.  Depending on their attitude values, they can also push the major groups to slowly polarize more than they would do without these minor clusters. For a long time in social science, minorities had reputed having no effect on majority groups. It has now changed; they appear as a source of creativity and interrogation, a sort of openness or alternative. Indeed, they avoid too much stability and often shake the public debate. Our model may account for such dynamics.
  
The second typical opinion evolution occurs when the rejection threshold is significantly higher than the attraction threshold. For these values, in these the simulations the rejection process is dominant. The initial attraction on the secondary dimension is weak. Indeed, as the attraction on the main dimension is low, many individuals stay far from each other. The polarization on this dimension begins before the groups have been really formed. In this case, people are very narrow-minded about what is important for them. Thus they form a lot of groups. Moreover, individuals want to be very different from individuals of other groups on the secondary dimension. They socially define themselves by differentiation to others. It results on constant fluctuations on the secondary dimension.

These fluctuations remain in-group fluctuations on the secondary dimension if the rejection threshold is not too high. Groups are less cohesive on the least important dimension. Individuals continuously define themselves on this dimension by differentiation to the other groups. However, they define themselves as a member of their group on the important dimension. 

When the rejection threshold becomes even larger, individuals remain in a continuous indecision and always fluctuate without being able to form a group on this secondary dimension. This can remind political regimes with a lot of small parties which are subject to frequent tactical changes of positions to differentiate from each other. However, this particular opinion evolution does not fit any observation from the experiments we took as a source of inspiration. A deeper investigation in the social psychological work would help to determine if it can be related to precise observations.

Another effect of a very large rejection threshold is the creation of extremist groups. The cohesiveness and the stability of these groups depend on the attraction threshold. Individuals composing these groups fluctuate a lot on the secondary dimension when the attraction threshold is low, as mentioned previously. However, when the attraction threshold is large, these extremist groups are stable and it is the centrist groups which is less cohesive. This latter situation sounds more realistic. In the political domain, the groups of extremists are generally cohesive even on a question which does not define their groups, whereas, the more centrist groups are more likely to vary on questions which are not group-relevant. 

Complementary investigations would be useful to check the robustness of these conclusions:
\begin{itemize}
	\item The initial distribution of attitudes has an impact on the stationary state configuration, and using the uniform distribution is not the most realistic hypothesis. Testing other rules for initialization could be useful.
	\item We should also vary the speed of the attitude move (parameter $\mu$), since we noticed that this parameter can have a strong impact on our first model with rejection [34] but also because it is probably able to suppress the minor clusters as suggested by [37].
	\item It would be worth considering a distribution of values for the thresholds instead of considering that all individual share the same values. 
	\item We should study the model with more than two attitudes and determine the impact of a selective discussion (an individual has to choose what it wants to discuss).
\end{itemize}

\end{document}